\documentclass[runningheads]{llncs}
\usepackage[T1]{fontenc}
\usepackage{graphicx}
\usepackage[backend=biber,
  style=lncs
]{biblatex}

\usepackage{graphicx}
\usepackage{hyperref}
\usepackage{amsmath}
\usepackage{amsfonts}
\usepackage{subcaption}
\usepackage{placeins}
\usepackage{array}
\usepackage[capitalise]{cleveref}
\usepackage{wrapfig,booktabs}
\usepackage{multirow}
\usepackage{tikz}
\usetikzlibrary{decorations.pathreplacing}
\usepackage{adjustbox}
\usepackage{makecell}
 
\usepackage{algorithm}
\usepackage{algpseudocode}

\algrenewcommand\algorithmicindent{0.5em}%

\usepackage{color}

\usepackage{xcolor}
\definecolor{darkblue}{HTML}{004C99}
\definecolor{lightblue}{HTML}{009999}

\DeclareMathOperator{\N}{\mathbb{N}}
\DeclareMathOperator{\R}{\mathbb{R}}
\DeclareMathOperator{\E}{\mathbb{E}}
\DeclareMathOperator{\DKL}{D_{\text{KL}}}

\DeclareMathOperator*{\argmin}{\arg\!\min}
\begin{document}
\title{An Active Inference Model of Mouse Point-and-Click Behaviour}
\author{Markus Klar\inst{1}\orcidID{0000-0003-2445-152X} \and
Sebastian Stein\inst{1}\orcidID{0000-0003-1828-4008} \and
Fraser Paterson\inst{1}\orcidID{0009-0003-4392-7092} \and
John H. Williamson\inst{1}\orcidID{0000-0001-8085-7853} \and
Roderick Murray-Smith\inst{1}\orcidID{0000-0003-4228-7962} }
\authorrunning{M. Klar et al.}
\institute{University of Glasgow, Glasgow, Scotland, UK}
\maketitle              %
\begin{abstract}

We explore the use of Active Inference (AIF) as a computational user model for spatial pointing, a key problem in Human-Computer Interaction (HCI).
We present an AIF agent with continuous state, action, and observation spaces, performing one-dimensional mouse pointing and clicking.
We use a simple underlying dynamic system to model the mouse cursor dynamics with realistic perceptual delay.
In contrast to previous optimal feedback control-based models, the agent's actions are selected by minimizing Expected Free Energy, solely based on preference distributions over percepts, such as observing clicking a button correctly.
Our results show that the agent creates plausible pointing movements and clicks when the cursor is over the target, with similar end-point variance to human users. In contrast to other models of pointing, we incorporate fully probabilistic, predictive delay compensation into the agent.
The agent shows distinct behaviour for differing target difficulties without the need to retune system parameters, as done in other approaches.
We discuss the simulation results and emphasize the challenges in identifying the correct configuration of an AIF agent interacting with continuous systems.

\end{abstract}

\section{Introduction}
Pointing is a fundamental interaction technique and \textit{Human-Computer Interaction (HCI)} has a long history of studying human pointing behaviour \cite{myers2024pick}. Pointing is a way of communicating information through an interface by manipulating a user-controlled cursor into spatial partitions corresponding to actions; typically, these are discrete actions like selecting a menu item. Modelling pointing behaviour is crucial in engineering interactions well-suited to human capabilities, or in predicting the behaviour of users in using an existing interface without real users in the loop. As a prevalent and tractable interaction problem, it has attracted substantial attention in the HCI literature.  HCI has historically focused on high-level descriptive models of pointing using summary statistics. Fitts' Law~\cite{fitts1954information,fitts1964information,mackenzie_fitts_2018} is the classic model, and relates the expected time taken to acquire a target (with a certain tolerated error rate) to the size and distance of the target from the cursor origin. While such models are effective in explaining movement time and are simple to apply, they do not attempt to model the detailed process of target acquisition itself, and do not consider click dynamics. 
Insights into complete cursor trajectories would enable more precise prediction of user intentions and the development of improved assistive tools.

One way to improve this understanding is to build a fine-grained, computational, generative model of the human user that synthesizes the entire target acquisition process. Approaches to model whole pointing trajectories have been introduced, e.g. in ~\cite{seungwon2021point,MulOulMur17,fischer_optimal_2022,martin2021intermittent,klar2023mpc}, often with models derived from a control-theoretic perspective, such as the Minimum Jerk model~\cite{flash1985coordination} or the E-LQG~\cite{todorov2005generalized}.
Other approaches have applied machine learning to derive policies that emulate human behaviour from data \cite{cheema2020midair,dalsgaard2021pointing,ikkala2022uitb}. These classical models typically rely on the design of cost or reward functions, which are often defined by a combination of the distance between the cursor and the target and the control effort required to reach it. They are also typically deterministic in that they do not model or represent uncertainty explicitly, some even requiring a predefined movement time. Furthermore, in contrast to our approach, these focus on cursor movement only and clicking is usually not included, or seen as a separate task.

\begin{figure}
    \centering
    \begin{subfigure}[t]{0.5\linewidth}
        \centering  
        \includegraphics[height=0.2\paperheight]{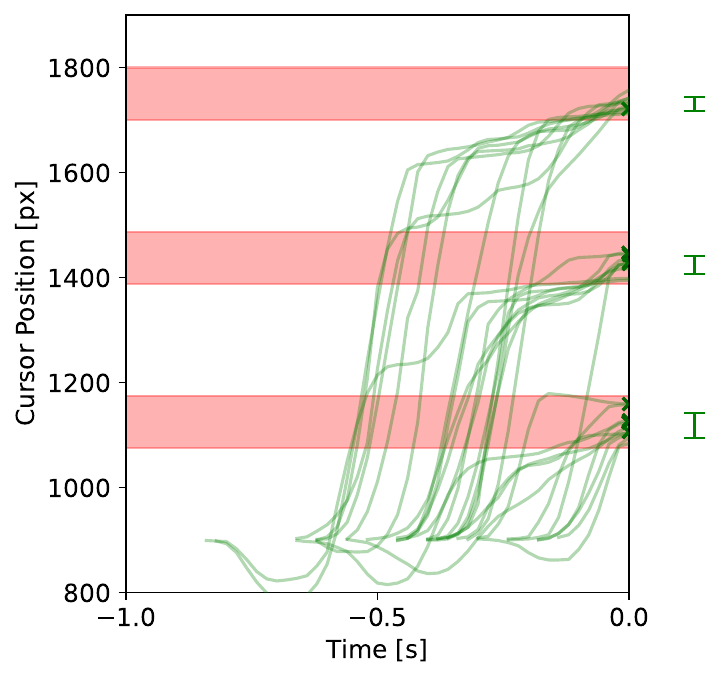}
    \end{subfigure}%
    ~ 
    \begin{subfigure}[t]{0.5\linewidth}
        \centering  
        \includegraphics[height=0.2\paperheight]{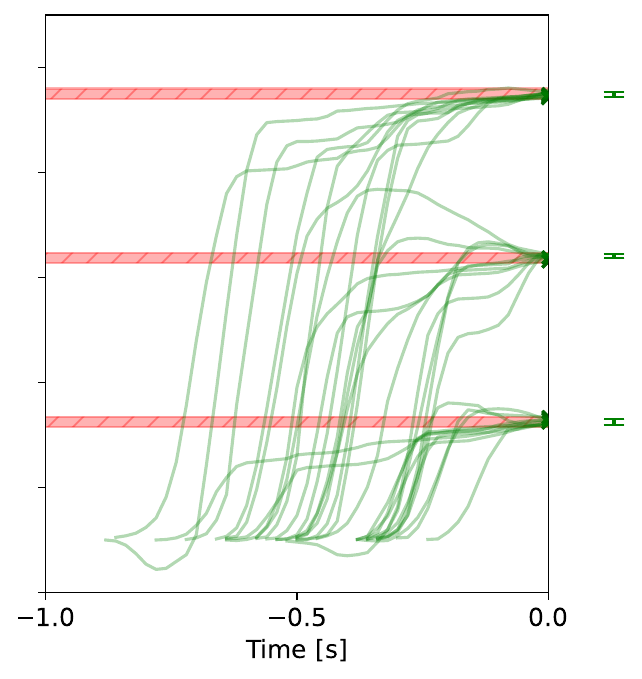}
    \end{subfigure}
    \caption{Mouse cursor trajectories (green lines) and clicks (green crosses)  of one human user pointing and clicking at targets (red) with different distance (850px, 537px, 225px) and different widths (left: 100px, right: 20px). $t=0$ indicates the moment of a correct click. The error bars show the end-point standard deviations.}
    \label{fig:user-trajectories-intro}
\end{figure}

We propose the application of active inference (AIF)~\cite{parr_active_2022} to create a pointing model that is fully probabilistic, predictive and formulates the problem in terms of preferences over observations, rather than rewards. Different behaviour for difficult targets, which occurs in real user data (see~\cref{fig:user-trajectories-intro}), should emerge from greater uncertainty in the target acquisition rather than changed model and reward parameters -- which is often the case in the classical models of pointing mentioned above. 
As an underlying model, we make use of the well-understood Second-Order Lag model which is commonly used to model human mouse pointing~\cite{SheFer74,MulOulMur17,fischer_optimal_2022}. 
Using such a simple system to represent the human motor control system allows us to focus on the AIF agent's inference and control.
We additionally apply perceptual noise and delay in order to achieve a more human-like behaviour~\cite{perrinet2014active,priorelli2023flexible}. 
AIF is a widely-applicable model of the behaviour of organisms that unifies Bayesian models of perception with a probabilistic model for action under uncertainty. It postulates behaviour as a process of reducing surprise to drive the environment into preferred configurations and minimizing uncertainty that might, in the future, lead to disfavoured states.

Recently there has been interest in bringing active inference concepts into the study and practice of HCI, see reviews such as \cite{MurWilSte24} and \cite{vertegaal2025}. Stein et al. \cite{SteWilMur24} explored ordinal selection with AIF, i.e., selecting the target from a set of discrete elements, but modelling pointing with AIF has not been explored. In the context of pointing, AIF offers a fundamentally different perspective. An AIF formulation frames pointing as a continuous process of minimizing prediction error by continuously updating internal beliefs based on sensory input.  Active inference defines the goals of a (simulated) user in terms of a distribution over desired observations.  For example, rather than moving the cursor exactly to the centre of the target, the user might be satisfied as long as there is a high chance of clicking the button correctly. However, AIF has typically been applied in discrete contexts or in restricted forms. Modelling pointing, at the level of dynamic trajectory generation, requires a continuous representation of actions (such as potentials to be propagated to a simulated motor system), states, and observations (such as cursor position). We propose an AIF model that is capable of modelling fine-grained pointing behaviour in this continuous setting, and which generalises to different tasks and environments. 

\ \\
In summary, our contributions are:
\begin{itemize}
    \item A continuous active inference formulation of point and click behaviour which does not rely on arbitrary discretisations, appropriately models perceptual noise and delay, and provides probabilistic predictions of cursor trajectories.
    \item An empirical study of a classical 1D pointing task with our active inference model, including a comparison of pointing trajectories with behaviour observed in a user study.
\end{itemize}

\section{Active Inference for Continuous Dynamic Systems with Observation Noise and Perceptual Delay}
Control of a human limb is an example of the general problem of control of a dynamic system. \cref{sec:AIFequations} describes the notation for such continuous system dynamics, how the agent constructs a belief about the states, dynamics and noise, as well as how it updates them. Specific adaptations of the standard general active inference approach in our model include:
\begin{itemize}
\item{\bf System Dynamics:} The general equations are described in \cref{sec:syst_dyn}. In the context of modelling a human pointing with a mouse, the system state may be a combination of the cursor's position and velocity and whether the mouse button is being clicked. 
The observation reflects the user's visual, proprioceptive, or haptic perceptions, which are noisy by nature.
\item{\bf Probabilistic Beliefs of AIF Agent} are constructed about the states, dynamics and  noise:
1. Belief about system state 
$Q^s := \mathcal{N}(\mu^s, \Sigma^s)$.
2. Belief about parameters of the system dynamics $Q^\theta := \mathcal{N}(\mu^\theta, \Sigma^\theta)$.
3. Belief about noise in observations $Q^p := \exp (\mathcal{N}(\mu^p, \Sigma^p))$. %
Each of these beliefs is represented by the mean and covariance matrix of a (multivariate) Gaussian distribution.
Since the belief about the noise $Q^p$ describes the standard deviation of the different noise factors and thus needs to be positive at all times, we choose a log-normal distribution for robustness.
\item{\bf Belief Update mechanisms} are mostly standard, and are described in \cref{sec:belief-updates}. To efficiently update the agent's belief after performing an action, we use an Unscented Kalman filter (UKF) which can be used to propagate Gaussian random variables through system dynamics~\cite{wan2000unscented,julier2004unscented}. 
\item{\bf Action Selection} is based on minimising Expected Free Energy (EFE) as described in \cref{sec:action-selection}. Since these calculations require a lot of computation time, especially when the horizon $N$ is large, the evaluation in this work will exclude simulations with information gain.
\item{\bf Perceptual Delay} is modelled by constructing a belief about $s(t-\tau)$ for a given perceptual delay of $\tau \in \N$ timesteps and updating this belief using the observation $o(t-\tau)$ obtained in step $t$. Before action selection, the agent uses its generative model and the last actions to create a temporary belief about $s(t)$ by sequential updates. Additionally, we let the agent know about the exact target location only after the initial delay (see~\cref{sec:belief-priors-and-delay}). 
\end{itemize}

\section{1D Mouse Pointing with Active Inference}\label{sec:experiment}
We now demonstrate how AIF can be applied to simulate a human user performing a 1D-mouse pointing task. 
In such a task, the user is asked to position a mouse cursor within a one-dimensional target whose size and distance is varied systematically.
A detailed description about the pointing task is provided in~\cref{app:experiment-details}.
We simulate single trials which terminate as soon as the button is clicked successfully or after two seconds.
Due to numerical constraints, both the canvas and the targets are scaled down by a factor of 1000. Furthermore, the initial position is recalculated to be centred at the origin.
All shown results are however rescaled to the original size.
We apply our model to investigate whether a cursor controlled with AIF:
1. allows us to model mouse clicking based on probabilistic beliefs about the system and preference distributions;
2. shows similar variability to observed human movements;
 and 3. can model the variety of behavioural strategies shown for different target difficulties without separate models for each. %

\begin{figure}
\includegraphics[width=\linewidth]{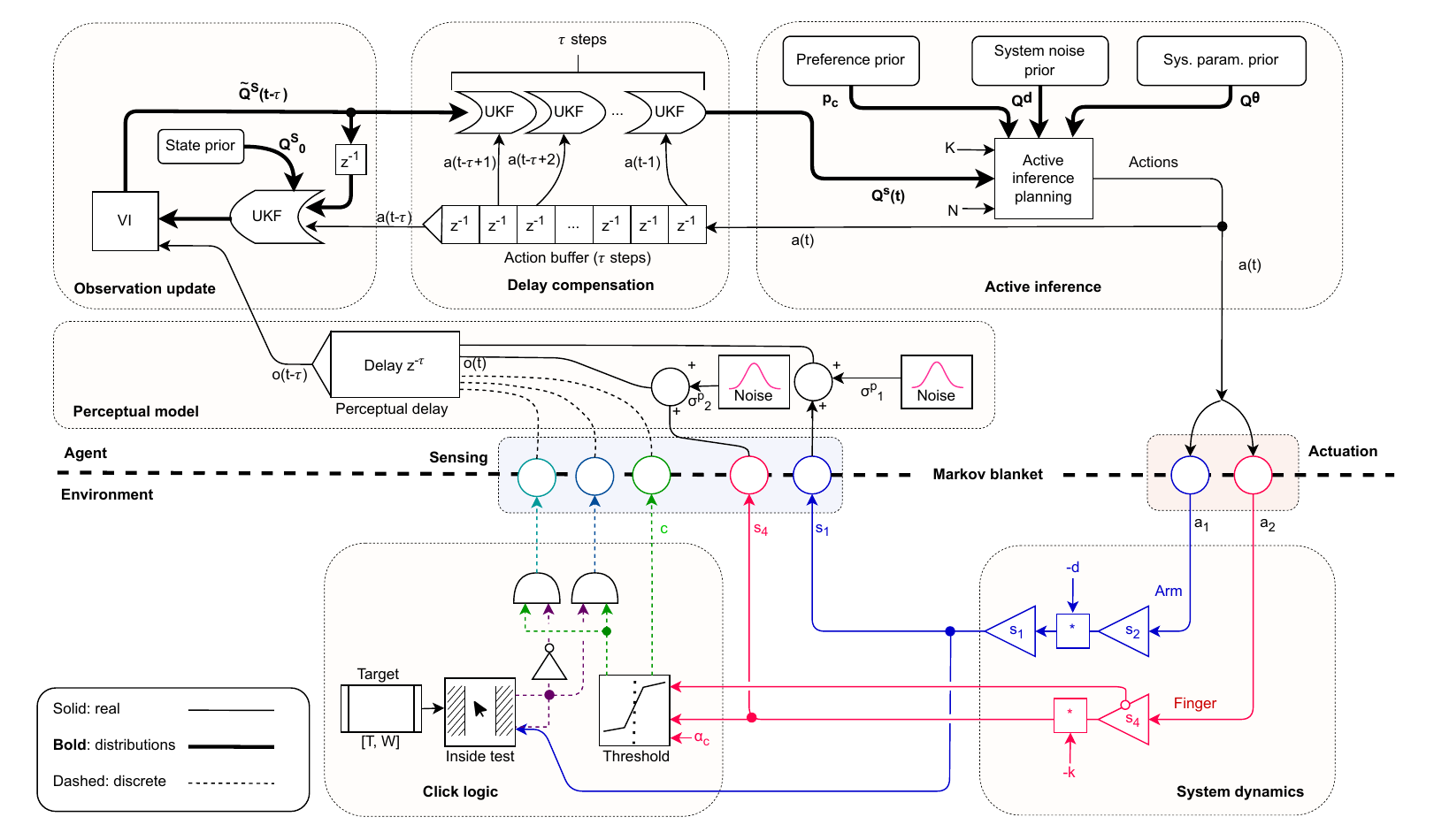}
\caption{A block diagram showing the interaction loop of the agent (upper) and the environment (lower). The agent includes a perceptual model, predictive delay compensation and min-EFE action planning. The simulated environment consists of a first and second-order lag modelling cursor and finger dynamics, along with click handling logic. See \cref{alg:algorithm} for full details.}\label{fig:block_diagram}
\end{figure}

\noindent
The complete interaction loop of the agent with the system is depicted as a block diagram in~\cref{fig:block_diagram} and provided as pseudo code in~\cref{app:algorithm}.

\subsection{The Generative Process and Model}\label{sec:genp-mouse}%
To simulate human pointing, we need to build a model for the pointing technique and interface.
In Active Inference terms, this means that we construct a generative process that plays the role of the `real world'.
The agent can interact with the process via a Markov Blanket which also defines what the agent observes. 
The agent also constructs a generative model of the world, which can be different from the real world.
We decided to keep the system dynamics identical, however, the generative model is governed by the agent's belief about system parameters (which may be wrong or uncertain).
Our model has two main components, the arm dynamics and the finger dynamics, and an additional logic for mouse clicks. 

\textbf{\textcolor{blue}{Arm Dynamics}} As a basis for our simulated arm dynamics we choose a Second-Order Lag model~\cite{SheFer74,MulOulMur17,fischer_optimal_2022}. 
The state of this system is given by the cursor position $s_1$ and velocity $s_2$. 
The agent controls the acceleration $a_1$ that is applied to the cursor.
When moving a computer mouse over a desk in a reasonable area, there is no force pulling the mouse to a certain stable position. 
Therefore we decided to set the cursor's stiffness to zero. 
However, damping is still applied with a factor of $d > 0$.

\textbf{\textcolor{red}{Finger Dynamics}} To model the index finger pressing the left mouse button, we use a simple First-Order Lag system~\cite{SheFer74} with stiffness $k >0$.
A click is only initiated when the displacement applied to the mouse button exceeds a threshold, requiring the agent to release the button before being able to click again.
Therefore, both the previous button displacement $s_3$ and the current button displacement $s_4$ must be present in the system state.
The agent can increase or decrease this force with a second action $a_2$.

After time-discretisation with $\Delta t = 0.02$, the combined system dynamics which take state $s$ and action $a$ and obtain the new state are therefore given by
\begin{equation}\label{eq:1d-mouse-pointing-dynamics}
f_\theta : \mathbb{R}^4 \times \mathbb{R}^2 \to \mathbb{R}^4,\quad 
f_\theta (s, a) = 
\begin{bmatrix}
\textcolor{blue}{s_1 + \Delta t\, s_2} \\
\textcolor{blue}{s_2 - \Delta t\, d\, s_2 + \Delta t\, a_1} \\
\textcolor{red}{s_4}\\
\textcolor{red}{s_4 - \Delta t\, k\, s_4 + \Delta t\, a_2}
\end{bmatrix},\quad
\end{equation}
where $d > 0$ is the damping parameter of the cursor and $k >0$ is the stiffness parameter of the finger.
The observation function is given by
\begin{eqnarray}\label{eq:1d-mouse-pointing-output}
g_\theta : \mathbb{R}^4 \to \mathbb{R}^5,\quad 
g_\theta(s) = 
&\begin{bmatrix}
\textcolor{blue}{s_1}  \\
\textcolor{red}{s_4} \\
\textcolor{green}{c} \\
\textcolor{darkblue}{c \land \left(| s_1 - T | \leq \frac{W}{2} \right)} \\
\textcolor{lightblue}{c \land \left(| s_1 - T | \leq \frac{W}{2} \right) - c} 
\end{bmatrix},\\
c &= (s_3 < \alpha_c) \land (s_4 > \alpha_c)
\end{eqnarray}
and contains the cursor position, the current button displacement, whether the mouse has been clicked, whether the button has been hit, and whether the button was missed.
The parameters $T$ and width $W > 0$ are the target centre and width, respectively, and $\alpha_c > 0$ is the clicking threshold. 
Thus, both functions depend on the system parameters $\theta = \left[d, k, T, W, \alpha_c \right]$.

\textbf{Click logic}
The observation of clicks works as follows.
If the previous button displacement was below and the current button displacement is above the threshold, a click is issued and $c = 1$. 
In any other case $c = 0$.\footnote{Following the convention that \texttt{True} is represented as $1$ and \texttt{False} as $0$.}
If, additionally, the cursor position is inside the target, i.e., $\left(| s_1 - T | \leq \frac{W}{2} \right)$, the button is clicked successfully, i.e., $g(s)_4 = 1$ (otherwise $g(s)_4 = 0$).
The last entry of the output represents a negative feedback for clicking outside the button, such as a beep tone commonly used in studies~\cite{MulOulMur17}.
As such, $g(s)_5 = -1$ when a misclick happens (otherwise $g(s)_5 = 0$).
We intentionally add this observation to allow the agent to have a specific preference of \textit{not} observing a misclick.

Since human perception is inherently subject to noise, we add Gaussian noise to the agent's observations of cursor position and button displacement. %
We assume that the other components of output are observed without any disturbances, leading to an observation of $o = \left[o_1, o_2, g(s)_3, g(s)_4, g(s)_5\right]$ with $o_1 \sim \mathcal{N}\left(g(s)_1,\sigma^p_{1}\right)$ and $o_2 \sim \mathcal{N}\left(g(s)_2,\sigma^p_{2}\right)$. %
Except otherwise noted, we use the parameters shown in~\cref{app:parameters}.

\subsection{Belief Priors and Perceptual Delay}\label{sec:belief-priors-and-delay}
We explore agents that are supposedly well-trained in the pointing task, i.e., the means of the prior belief about the system parameters $Q^\theta$ are close to the real values while the uncertainty is small.
However, to model the initial delay of perceiving the target, we initialise the agent's belief about the target location to be centred at the initial position and have a high uncertainty, i.e., the agent believes that the target could appear anywhere on the display.
In summary, the agent's initial belief $Q^\theta$ is given by
\begin{equation} %
\mu^\theta = \left[d, k, 0.0, 0.03, \alpha_c\right], \; \Sigma^\theta = \text{diag}\left(\left[0.2, 0.2, 0.9, 0.02, 10^{-6}\right]^2\right).
\end{equation}
After the delay period, we update the belief to include the correct target position $T$ and width $W$ with low uncertainty, that is,
\begin{equation}
    \mu^\theta = \left[d, k, \mathbf{T}, \mathbf{W}, \alpha_c\right], \; \Sigma^\theta = \text{diag}\left(\left[0.2, 0.2, \mathbf{10^{-6}}, \mathbf{10^{-6}}, 10^{-6}\right]^2\right).
\end{equation}
Note that we keep some uncertainty about the damping and stiffness. 
This is due to the fact that an agent with pristine knowledge about all system parameters would not need to infer the system state from noisy observations. 
Instead, its internal model of the world would allow for perfect prediction of the cursor trajectory and clicking, resulting in super-human performance.
However, to keep the sources of uncertainty in check, we provide the agent with full knowledge about the amount of observation noise, i.e. $Q^p$ is given by 
\begin{equation}
\mu^p = \left[\sigma^p_{1},\sigma^p_{2}\right] ,\; \Sigma^p = 10^{-8} I_2 .
\end{equation}

In future work, $Q^\theta$ and $Q^p$ might also be inferred by interacting with the system over a longer time period.

The agent starts with a reasonable initial belief about the system state, that is
\begin{align}%
\mu^s = \left[0.0, 0.0, 0.0, 0.0\right], \;
\Sigma^s = \text{diag}([0.001, 0.0001, 0.00005, 0.00005]^2) \ .
\end{align}
We address the time a perceived signal takes to be processed by a human as follows.
We delay observations that the agent received from the system after applying an action by $\tau=5$ timesteps (=100ms).
To compensate for this perceptual delay, the agent keeps a belief about the state before the delay, $Q^s(t-\tau)$ (instead of $t$). 
Prior to planning its next action, the agent then predicts the system state at time $t$ using its internal model.
This is achieved by sequentially updating a temporary copy $\tilde{Q}^s(t-\tau)$ of $Q^s(t-\tau)$ by using actions applied during the delay $a(t+i), i=-\tau,\dots,-1$, until arriving at $\tilde{Q}^s(t)$ (see~\cref{sec:belief-updates}, 1.). 
After planning and performing the next action, the agent uses the ``old'' action $a(t-\tau)$ and, if available, delayed observation from $o(t-\tau)$ to obtain the belief $Q^s(t-\tau+1)$.
This belief then serves as an initial point for the next planning step.

\subsection{Preference Prior}
The agent's goal is to control the generative process described in~\cref{sec:genp-mouse} in the same way a human does, i.e., such that the target button is clicked as accurately and fast as possible.
To model this in a natural way with Active Inference, we define the agent's preference distribution as a marginal distribution on the observations of the cursor position $o_1$, button click $o_4$, and misclick $o_5$,
\begin{equation}\label{eq:preference-distribution}
    P^c_{o_1,o_4,o_5} = \mathcal{N}\left([T, 1.0, 0.0],
    \text{diag}\left(\left[ 0.01, 0.01, 0.001\right]^2\right)
    \right).
\end{equation}
The mean of this distribution is centred around the cursor being at the target's centre, the button being successfully clicked, and not getting negative feedback on a misclick.
Adding the additional preference of not observing a misclick ensures that the agent only clicks when it is certain enough that the cursor is on the button.
Scaling the diagonal values of the covariance matrix allows creating agents that have different focus on speed versus accuracy, a classical trade-off that is present in human aimed movements~\cite{fitts1954information, schmidt1979motor, guiard2015mathematical}.

The parameters used in the results are manually fitted to the data of one user performing the pointing task described in~\cref{sec:data-acquisiton} and more details about the fitting process can be found in~\cref{app:parameters}. 
We additionally present the effect of changing individual parameters in~\cref{app:add_results}.
It is important to mention that we -- other than previous optimal control based approaches -- use the same parameter set for all targets. 
The qualitatively different behaviours emerge from the properties of the AIF agent instead.

\section{Results}
We investigate whether the AIF agent is able to produce movements which are similar to human-controlled movements, despite the simplicity of the underlying model.
In particular, we are interested in the amount of variability produced by the agent compared to a human performing the same task.
\vspace{-0.3cm}
\subsection{AIF Agent Performs Pointing Task}
\begin{figure}
    \centering
    \begin{subfigure}[t]{0.5\linewidth}
    \centering
        \includegraphics[height=.22\paperheight]{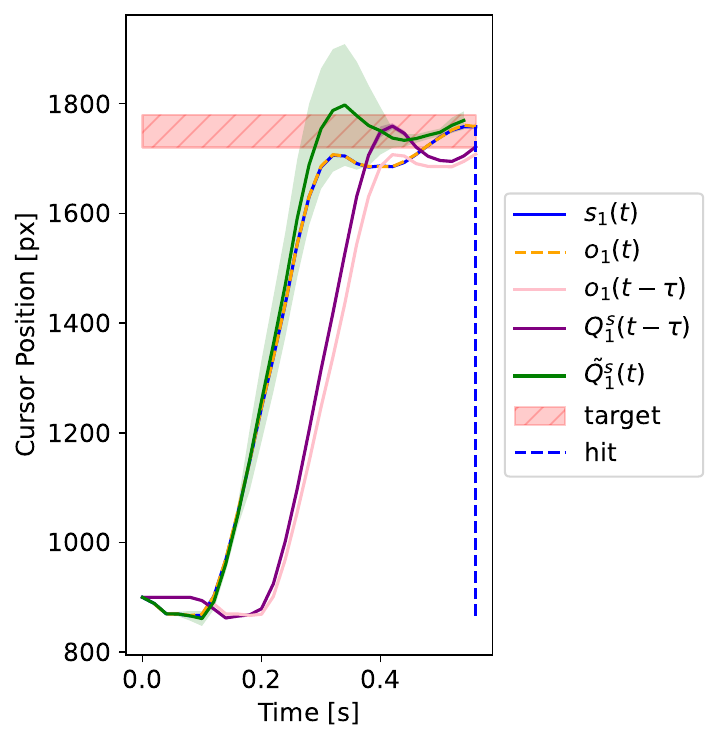}
    \end{subfigure}%
    \begin{subfigure}[t]{0.5\linewidth}
    \centering
        \includegraphics[height=.22\paperheight]{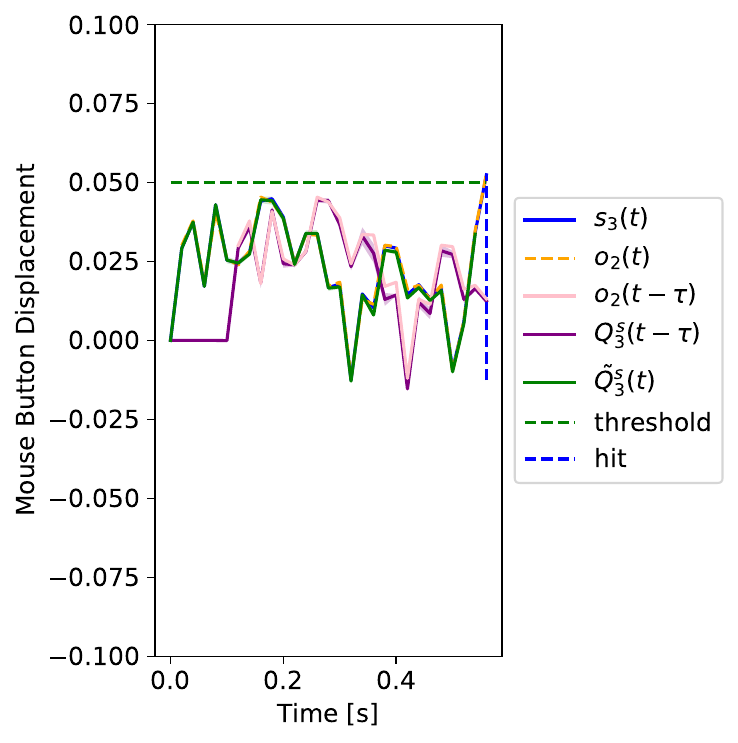}
    \end{subfigure}%
    \caption{Mouse cursor and button displacement for one trial of the AIF agent from the start at 900 to $T=1750$, $W =60$px. The plot shows the real state $s(t)$ (blue), the agent's belief about $s(t-\tau)$ at time $t$ (purple) and the prediction of $s(t)$ used for planning (green). The ribbon displays +/- 3 standard deviations.}%
    \label{fig:example_pointing}
\end{figure}
\noindent
\cref{fig:example_pointing} shows a time-series view of how the AIF agent performs the pointing task.
Before obtaining knowledge about the target position (before $0.1$s), the agent keeps the cursor close to the centre.
During the fast surge movement (an initial ballistic phase of a mouse movement, \cite{MulOulMur17}), the agent's perception of the cursor position (left plot) is subject to increased uncertainty due to the larger impact of the perceptual delay. %
After slowing down close to the target, the agent regains a more precise belief about the cursor position and ultimately manages to click on the target.
The resulting corrective movements are commonly observed in human pointing.
In the right plot, the button displacement is represented.
The initiation of a click is prompted upon attaining a threshold of $\alpha_c = 0.05$.
Given that the preference distribution is shaped by the observation of not missing the button, the agent only clicks when it believes that the cursor is inside the target.

\vspace{-0.3cm}
\subsection{AIF Agent Follows Fitts' Law}
Fitts' Law describes a logarithmic relationship between the ratio of a target distance to target width and the time for humans to reliably acquire it. 
It is usually given in the Shannon form: $MT = a + b \cdot \log_2\left(1 + \frac{D}{W}\right)$, where $MT$ is the movement time, $a$ and $b$ are empirical constants, $D$ is the distance to the target, and $W$ is the width of the target along the axis of motion~\cite{fitts1954information,mackenzie_fitts_2018}.  The log ratio term $\log_2\left(1 + \frac{D}{W}\right)$ is called the \textit{index of difficulty (ID)} and is measured in bits; larger values indicate targets which are more difficult to acquire.

\begin{figure}
    \centering
    \begin{subfigure}[t]{0.5\linewidth}
        \centering  
        \includegraphics[height=0.15\paperheight]{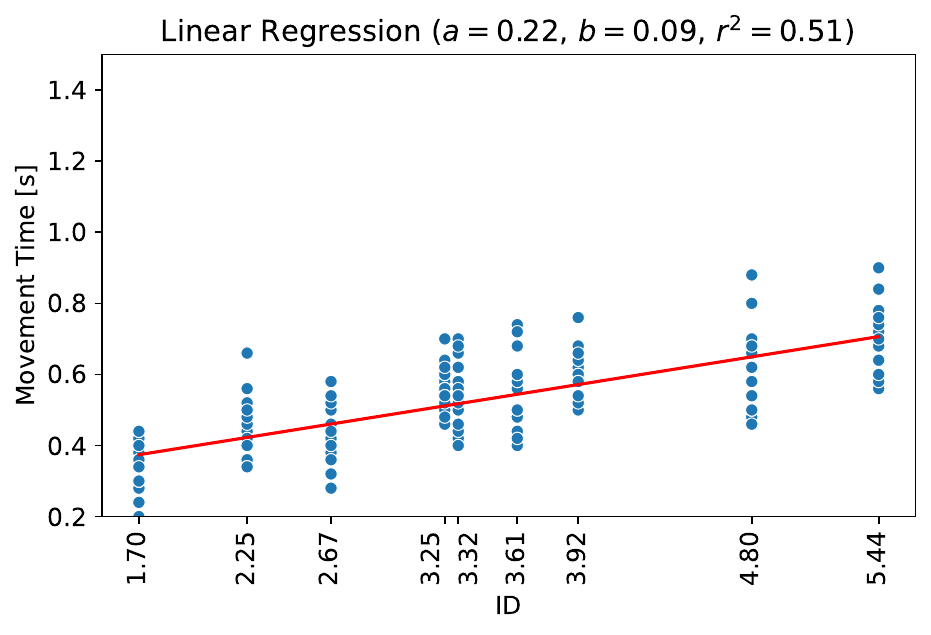}
        \caption{AIF Agent}
    \end{subfigure}%
    \begin{subfigure}[t]{0.5\linewidth}
        \centering  
        \includegraphics[height=0.15\paperheight]{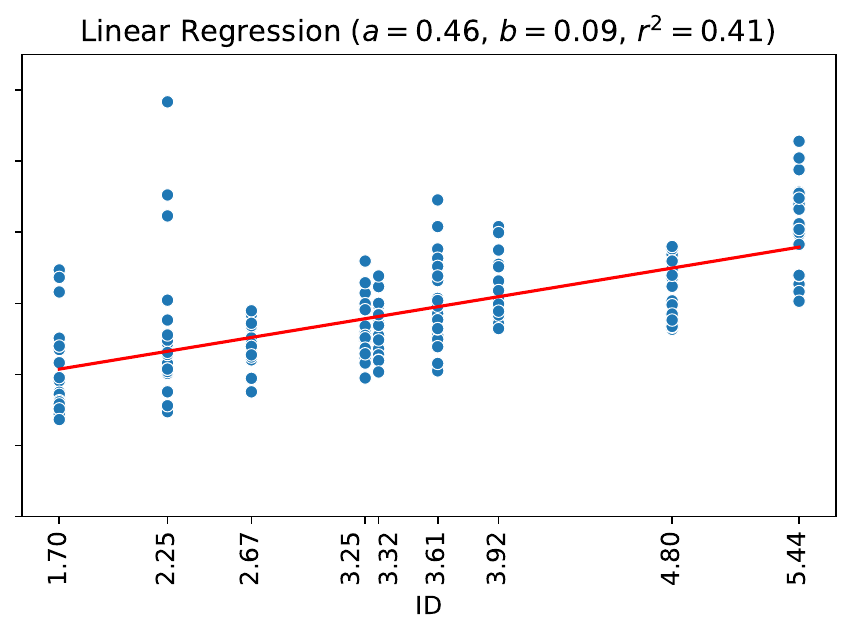}
        \caption{User}
    \end{subfigure}%
    \caption{Following Fitts' Law, human pointing movements show a linear relationship between movement time and the index of difficulty (ID). The simulation with AIF shows a similar trend. The data is fitted with a linear regression model. Simulation: $a = 0.22, b = 0.09, r^2 = 0.51$; User:  $a = 0.46, b = 0.09, r^2 = 0.41$. 
}
    \label{fig:fitts-law}
\end{figure}

If our model is a plausible model of human pointing, it should follow this relationship. 
If it is well-tuned for a specific configuration (such that we emulate a realistic human motor system with a particular pointing device), we should recover similar coefficients $a, b$ as in experimental data.
\cref{fig:fitts-law} shows the Fitts' Law evaluation  for the AIF agent and a human user side-by-side.
The AIF agent follows the characteristic pattern of Fitts' Law with an identical slope $b$ as the user, while having a shorter constant delay $a$.%

\vspace{-0.3cm}
\subsection{AIF Agent Shows Changing Behaviour for Different Targets}

\begin{figure}
    \centering
    \begin{subfigure}[t]{0.31\linewidth}
        \centering  
        \includegraphics[height=.25\paperheight]{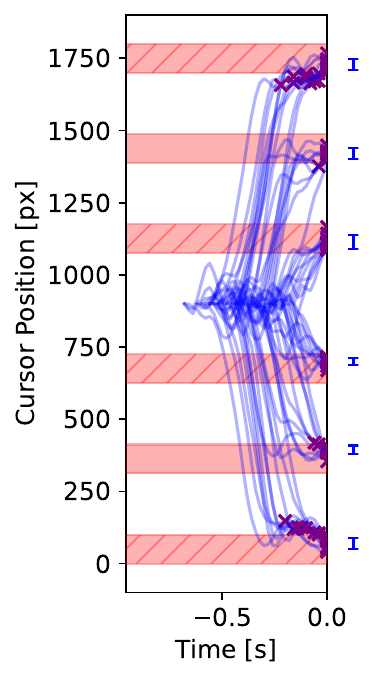}
    \end{subfigure}%
    \hfill
    \begin{subfigure}[t]{0.31\linewidth}
        \centering  
        \includegraphics[height=.25\paperheight]{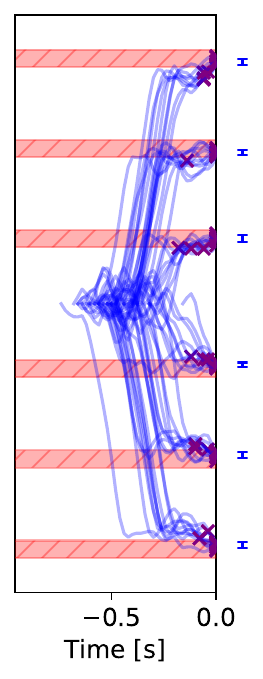}
    \end{subfigure}%
    \begin{subfigure}[t]{0.31\linewidth}
        \centering  
        \includegraphics[height=.25\paperheight]{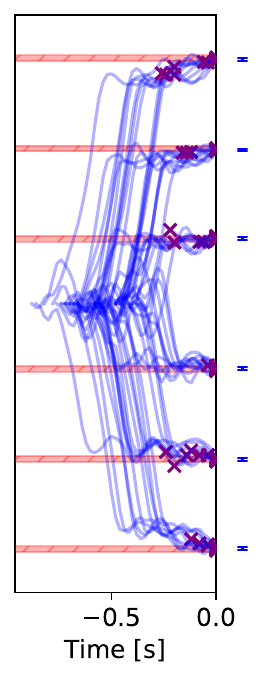}
    \end{subfigure}\\
    \begin{subfigure}[t]{0.31\linewidth}
        \centering  
        \includegraphics[height=.25\paperheight]{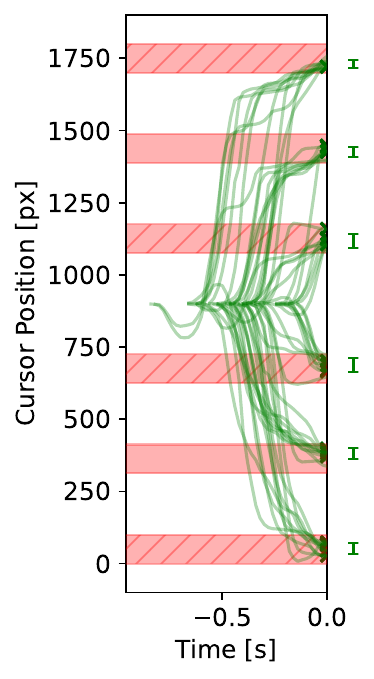}
    \end{subfigure}%
    \hfill
    \begin{subfigure}[t]{0.31\linewidth}
        \centering  
        \includegraphics[height=.25\paperheight]{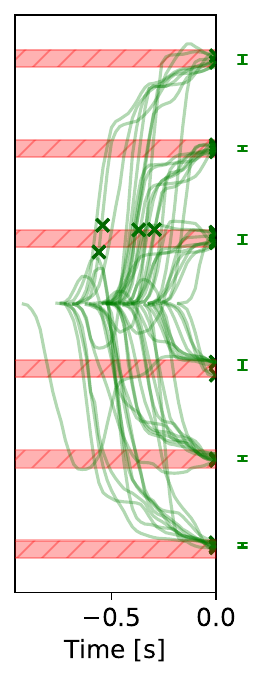}
    \end{subfigure}%
    \begin{subfigure}[t]{0.31\linewidth}
        \centering  
        \includegraphics[height=.25\paperheight]{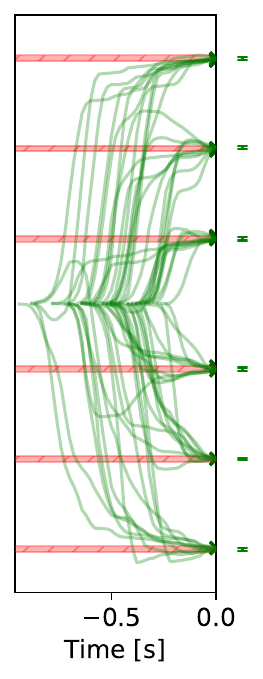}
    \end{subfigure}%
    \caption{Cursor trajectories for the AIF agent (upper row, blue) and a human user (lower row, green) for different target sizes (from left to right: 100px, 60px, 20px). The crosses mark mouse clicks, the red areas display the targets. $t=0$ indicates the moment of a correct click. The error bars show the end-point standard deviations. The averages for the different target sizes are: Agent: $(17.63, 10.08, 4.74)$, User: $(19.48, 13.02, 4.81)$.} %

    \label{fig:timeseries_aif}
\end{figure}

\cref{fig:timeseries_aif} shows the trajectories produced by the AIF agent and a human user, split up by target size.
As predicted by Fitts' Law, movements of both agent and human, are longer for more difficult targets, i.e., smaller targets (right plots) or targets with larger distance from the initial position at 900px. 
Similar to the human, if targets are smaller or further away, the agent-controlled cursor often shows one or multiple sub-movements, i.e., shorter corrective movements, after the surge phase.

Another characteristic of human behaviour is that it shows higher end-point variance for larger targets.
The end-point variance of the AIF agent's movements shows a similar trend, which is displayed via the error bars next to the trajectories in~\cref{fig:timeseries_aif}.
Similar to the user, for large targets (left plots in~\cref{fig:timeseries_aif}), the endpoints of the agent's movements are skewed towards the starting point, showing that the agent makes use of the target size.

\vspace{-0.3cm}

\section{Discussion}
\label{sec:discussion}

\textbf{Implications of the results}
The results indicate that a full active inference model of target acquisition and click actuation with appropriate perceptual delays leads to behaviours that are comparable to human performance. 
The gross behaviour, as quantified by the Fitts' law plots, indicates human-like pointing capacity, even without precise calibration. 
The AIF agent also shows realistic end-point variance and makes use of the target size. 
Unlike optimal control methods, where parameters have to be fit not just for each user but for each target, a single parameterisation leads to plausible pointing behaviour across targets spanning a range of indices of difficulty. 
The AIF agent's predictive models can compensate for the delayed observations with variability that appears to correspond well to human pointing in the presence of the well-known latency in the visual system. 
The combination of cursor movement and clicking enables interesting interactions, even in a simple model as ours, such as updating the belief about the cursor position by observing a click or misclick.

\textbf{Limitations}
Our model has a radically simplified representation of human perception and motor control as a second-order system with Gaussian-corrupted observations and fixed perceptual delay. Our parameterisations are hand-tuned and validated on the data of a single user. Our agent has a relatively short time horizon as a consequence of the computational cost of sample-based inference and the consequent effect of ``sample starvation'' on policy selection. 
These inference limitations have an implicit effect on the parameterisation of the agent -- for example, action selection is challenging if no sequence of actions in the time horizon brings the predicted state close to the ``correct click'' state. %
This can also lead to oscillating behaviour, as in~\cref{fig:timeseries_aif}, which is not as pronounced in humans.
The computational effort involved in the variational inference procedure and limit the amount of plans that can be evaluated.
However, our approach using UKF with Monte Carlo sampled rollouts inference makes our fully continuous formulation viable.

\textbf{Outlook}
Our model is a working AIF model of human pointing, but it required significant custom development. 
There is a clear need for software building blocks to make progress in AIF-based modelling of human interaction: reusable models and components appropriate for HCI tasks. 
We use a simple linear system as our forward model and a similarly simplified perceptual model with known delay. 
There remains opportunity to explore more authentic biomechanical models of the arm, or stronger perceptual models that, e.g. add uncertainty about the perceptual delay, properly account for motor delays, or include models of foveated vision. 
Although our model reproduces behaviour qualitatively similar to our test user, we have much to explore in calibrating parameters to better match human performance.
This could involve automatic optimisation of agent parameters to align with observed user data, and/or parameters informed directly by established biomechanical and perceptual data.
Optimisation could also be performed online, with an active inference agent learning system parameters during simulated interactions. 
We also see scope to develop more sophisticated metrics to establish what ``similar'' trajectories are, particularly when quantifying variability in dynamics. 
There are obvious extensions of our model to e.g. 2D or 3D pointing tasks, but these will likely require better inference, new datasets and stronger biomechanical and perceptual models.

\vspace{-0.4cm}
\subsection{Conclusion}
We presented an active inference simulation of human pointing and clicking behaviour with continuous states, actions, and observations. 
Despite its simple 2OL forward model our model generates plausible trajectories across a range of target sizes, and it includes a full model of actuation (``click'') planning and predictive delay compensation.

While AIF control of even this elementary system is not straightforward -- our agent has a profusion of parameters and significant inference challenges -- our findings suggest that AIF is a promising model for fine-grained interaction models. 
The inherently probabilistic nature allows for a greater variety of cursor trajectories than comparable methods, generating a wide variety of plausible behaviours. 
Our model is the first step towards active inference-based user modelling in HCI. 
Our results indicate that active inference is viable for user behavioural models and highlight many open challenges in building more sophisticated AIF-based synthetic users.

\noindent
The Python code used to create the simulations in this work can be found at \url{https://github.com/mkl4r/AIF-Pointing}.
\clearpage

\printbibliography

\appendix

\clearpage
\section{Active Inference for Continuous Dynamic Systems with Control and Observation Noise}\label{sec:AIFequations}

\subsection{System Dynamics}
\label{sec:syst_dyn}
We control a general dynamic system with continuous states, actions, and observations with active inference. 
We take a discrete-time approach, with a constant time step $\Delta t$.
A dynamic system is then described by functions 
\begin{eqnarray}\label{eq:sys_dyn}
s_{t+1} &=&f_\theta(s_t,a_t), \label{eq:dynamics}\\
o_t &=&g_\theta(s_t)+\epsilon_t, \label{eq:sys_out}
\end{eqnarray}
where (\ref{eq:dynamics}) maps the current system \textit{state} $s_t \in \R^n$ and an \textit{action} $a_t \in \R^m$ at timestep $t \in \N$ to the new system state $s_{t+1} \in \R^n$, given some system parameters $\theta$.
The system's output (\ref{eq:sys_out}) is given by a function $g_\theta: \R^n \mapsto \R^l$ and may be perturbed by constant and/or state-dependent Gaussian noise to obtain the agent's \textit{observation} $o_t \in \R^l$ at time step $t$.

\subsection{Belief Updates}\label{sec:belief-updates}
A human user will usually only have access to a limited set of system states that they observe via output devices such as displays. 
For instance, although we cannot perceive the velocity of a mouse cursor directly (as in a speedometer), it is still an important source of information to prevent overshooting a target button. 
The agent \textit{infers} this information by propagating its belief about the cursor velocity through its internal model, as explained below.

The agent has two ways to update its belief about the system state $Q^s_{t}$:
\noindent 1. After performing an action, the agent uses its internal model of the world to update its belief, resulting in a new belief $Q^s_{t+1}$ for timestep $t+1$. 
An efficient approach for this is to propagate the agent's belief through time using an Unscented Kalman filter (UKF)~\cite{julier2004unscented}, which allows the estimation of states under nonlinear transition functions with Gaussian belief distributions. 
It involves choosing sigma points in the combined space of $Q^s$ and $Q^\theta$, evaluating $f_\theta$ on these with the chosen action $a_t$ to obtain the corresponding $s_{t+1}$, and estimating the parameters of the updated belief distribution $Q^s_{t+1}$ directly from the $s_{t+1}$ using Maximum Likelihood.
Note that during this step we fold in the application of the action $a_t$ into the transition function that the UKF applies to make predictions of the future state under action $a_t$. 
This update usually increases uncertainty, since the agent is uncertain about the current state of the world as well as the dynamics.

\noindent 2. After a new observation, the agent updates its belief using variational inference.
This means, that the agent minimizes the variational free energy $F$ with respect to the belief distribution parameters and the noise belief,%
\begin{align*} \label{eq:belief_update_o}
\min\limits_{\hat{Q}^s} F\left(\hat{Q}^s, o_t\right) &= \DKL\left(Q^s \; \middle\| \; \hat{Q}^s\right) - \E_{s_t \sim \hat{Q}^s; \sigma_p \sim Q^p}\left[\ln P_o(o_t | s_t; \, \sigma_p)\right],
\end{align*}
\noindent where $Q^s$ is the variational belief \emph{prior} to making the new observation (but after having taken the last action), $\hat{Q}^s $ is the \emph{posterior} belief, and $\DKL$ the KL-divergence. 
The first term has an analytical solution for Gaussian belief distributions, whereas the second term requires sampling from $\hat{Q}^s$. 
For optimization via gradient descent, we use the reparameterization trick \cite{kingma2013auto} to backpropagate through $n^\text{VI}_s$ samples in each of the $k^{\text{VI}}$ update steps. 
Depending on the knowledge of observation noise, this step may reduce the agent's uncertainty.

\subsection{Action Selection}\label{sec:action-selection}
In every timestep $t$, the agent generates $K$ plans representing sequences of actions with a $N$-step planning horizon by sampling actions from uniform distributions, 
\begin{align*} 
\pi_j &= (a_{1j}, \dots, a_{Nj}), \forall j \in \{1, \dots, K\} \\ a_{ij} &\sim \mathcal{U}(\underline{a},\overline{a}),
\end{align*}
where $\underline{a},\overline{a}$ are the lower and upper boundaries for the actions taken by the agent.
These plans are evaluated with respect to the \emph{Expected Free Energy (EFE)}~\cite{parr_active_2022}, %
\begin{align*} 
G(\pi_j) &= \frac{1}{N}\sum_{i=1}^{N}G_{ij}(a_{ij}) \\
G_{ij}(a_{ij}) &= -\underbrace{\E_{Q^s_{ij}}\left[ \DKL\left(\hat{Q}^s_{ij} \; \middle\| \; Q^s_{ij}\right) \right]}_{\text{Information Gain}} - \underbrace{\E_{s \sim Q^s_{ij}; \sigma_p \sim Q^p; \tilde{o} \sim  \mathcal{N}(g(s), \sigma_p), }\left[ \ln P^c(\tilde{o}) \right]}_{\text{Pragmatic Value}},%
\end{align*}%
\noindent where $Q^s_{ij}$ are predicted via rollout, i.e. by successively updating the agent's belief after performing action $a_{ij}$ as described in~\cref{sec:belief-updates}.
Since these rollouts are predictions `in the agent's mind', the beliefs cannot be updated using observations here.\footnote{In discrete systems, Sophisticated Inference can be applied to walk the complete tree of possible actions and observations. However, even in cases where this is possible it is often too computationally expensive \cite{friston2021sophisticated}.}
In practice, it is often impossible to calculate these analytically, so we  approximate based on observations sampled from the current belief.
In case of perceptual delay, the initial belief distribution is obtained by applying UKF with the last performed actions as described in~\cref{sec:belief-priors-and-delay}. 

The \textit{information gain} is approximated by computing the KL-divergence between $Q^s_{ij}$ and different instances of $\hat{Q}^s_{ij}$.
This involves taking $n_s^\text{IG}$ samples $s \sim Q^s_{ij}$ and $\Xi_p \sim Q^p$ which are used to sample $n^\text{IG}_o$ observations from the generative model each, i.e. $\tilde{o} \sim \mathcal{N}(g_\theta(s), \Xi_p)$.
After performing the update after observation, as described in \cref{sec:belief-updates} for each of these, we obtain $n^\text{IG}=n^\text{IG}_s \cdot n^\text{IG}_o$ different instances of $\hat{Q}^s_{ij}$.

The \textit{pragmatic value} represents the bias encouraging the agent to take actions that result in preferred observations, which are characterized by the preference distribution $P^c$. 
As above, we sample $n_s^\text{PV}$ states from the current belief and sample $n^\text{PV}_o$ observations each to estimate the log-likelihood of these given the preference distribution.
The more likely these observations are, the better they and the plan leading to them are in the eyes of the agent.

The next action is chosen as the first action of the plan with minimal expected free energy, i.e. $a_{t+1} = a_{1j} \text{, where } j = \argmin_{j\in \{1,...,K\}} \; G(\pi_j)$.

\section{Implementation Details}

\subsection{Parameters}\label{app:parameters}
Our AIF agent has many parameters that can have an influence in the resulting trajectories. 
The following~\cref{tab:parameters} gives an overview of these parameters.
To find suitable parameters that lead to movements similar to those of a human, we proceed as follows.
\begin{itemize}
    \item We roughly estimate the necessary damping $d$ and maximal hand control $\overline{a}_1$ by comparing the 99\% percentile and maxima of accelerations and velocities observed in the study to those created by applying the maximal activation to the mouse cursor system~\cref{eq:1d-mouse-pointing-dynamics} for one second.
    \item We aim to set the stiffness of the finger dynamic such that the earliest click can not happen faster than observed in a user clicking as fast as possible, that is $\approx0.14$s per click. 
    However, this had to be relaxed slightly to ensure that the agent can click the button consistently. %
    \item We obtain the reaction time of the agent by computing the first timestep after the user exceeds an acceleration threshold of 10px/$\text{s}^2$, that is $\tau\approx0.1$s.
    \item We tune the preference distribution $P^C$ such that the resulting behaviour shows similar clicks and misclicks as humans.
\end{itemize}
The remaining parameters are tuned by visually comparing trajectories and the Fitts' Law results.
\begin{table}[]
    \centering
    \begin{adjustbox}{max width=\textwidth}
    \begin{tikzpicture}[remember picture]
    \node[inner sep=0pt] (table) {
    \begin{tabular}{l|c|c|l}
        \textbf{Variable} & \textbf{Domain} & \textbf{Value} & \textbf{Description} \\
        \hline\hline
        $d $ & $\R_{+}$ & $24.0$ & Cursor damping \\
        $k $ & $\R_{+}$ & $10.0$ & Button stiffness \\ 
        $T $ & $ \R$ & $*$ & Target position [px] \\ 
        $W $ & $\R_{+}$ & $*$ & Target width [px] \\
        $\alpha_c $ & $\R_{+}$ & $0.05$ & Click threshold \\
        $\Delta t $ & $\R_{+}$ & $0.02$ & Time step length [s] \\
        $\tau$ & $\N$ & $5$ & Perceptual delay [timesteps] \\
        $\underline{a}$ & $\R^2$ & $(-50.0,-1.0)$ & Control lower bounds \\
        $\overline{a}$ & $\R^2$ & $(50.0,1.0)$ & Control upper bounds \\
        $\Sigma_p $ & $ \R^{3}$ & $**$ & Std/Covariance of observation noise \\
        \hline
        $\mathcal{Q}^s $ & $ \R^{4}\times \R^{4\times 4}$ & $**$ & State prior \\  
        $\mathcal{Q}^\theta $ & $ \R^{5} \times \R^{5\times5}$  & $**$ & System parameters prior \\ 
        $\mathcal{Q}^d $ & $ \R^{3} \times \R^{3\times3}$ & $**$ & Noise prior \\ 
        \hline
        $N$ & $\N$ & $12$ & Prediction horizon [timesteps]  \\ 
        $K $ & $ \N$ & $3000$ & \# Plans \\
        $P^c $ & $ \R^3 \times \R^{3\times 3}$ & $**$ & Preference distribution \\
        \hline
        $k^{\text{VI}}$ & $ \N$ & $30$ & \# Optimization steps for VI\\ 
        $n^\text{VI}_s$ & $ \N $ & $300$ & \# State samples for VI\\
        $\eta^\text{VI} $ & $(0,1)$ & $3.0 \cdot 10^{-4}$ & Learning rate for VI\\
        $n^\text{IG}_s$ & $ \N $ &$X$ & \# State samples for Information Gain \\
        $n^\text{IG}_o$ & $ \N $ & $X$ & \# Observation samples for Information Gain \\
        $n^\text{PV}_s$ & $ \N $ & $50$ & \# State samples for Pragmatic value \\
        $n^\text{PV}_o$ & $ \N $ & $3$ & \# Observation samples for Pragmatic value
    \end{tabular}
    };

    \draw[decorate,decoration={brace,amplitude=10pt,mirror}, thick]
        ([yshift=5pt]table.south west)++(-0.4,11.4) -- ++(0,-5)
        node[midway,xshift=-1.0cm,rotate=90,anchor=center] {System};
    \draw[decorate,decoration={brace,amplitude=10pt,mirror}, thick]
        ([yshift=5pt]table.south west)++(-0.4,6.4) -- ++(0,-1.5)
        node[midway,xshift=-1.0cm,rotate=90,anchor=center] {Belief};
    \draw[decorate,decoration={brace,amplitude=10pt,mirror}, thick]
        ([yshift=5pt]table.south west)++(-0.4,4.9) -- ++(0,-1.5)
        node[midway,xshift=-1.0cm,rotate=90,anchor=center] {Control};
    \draw[decorate,decoration={brace,amplitude=10pt,mirror}, thick]
        ([yshift=5pt]table.south west)++(-0.4,3.4) -- ++(0,-3.5)
        node[midway,xshift=-1.0cm,rotate=90,anchor=center] {Hyper parameters};

    \end{tikzpicture}
    \end{adjustbox}
    \caption{List of Parameters. The value column shows the commonly used value, if not otherwise stated in the paper. $*$: Value is given by the task; $**$: Value is explicitly explained in~\cref{sec:experiment}; X: Value is not used for the results. For details see~\cref{sec:AIFequations}.}
    \label{tab:parameters}
\end{table}

\subsection{The Complete Algorithm}\label{app:algorithm}
\cref{alg:algorithm} shows the pseudo code for the complete interaction loop of an AIF agent with the mouse model. 
Prior to planning the next action, the agent compensates for the perceptual delay by estimating the system state at time $t$.
This is achieved by sequentially updating the temporary copy $\tilde{Q}^s$ using the previously applied actions (for-loop in planning phase) and the agent's internal model of the system. 
After planning and performing an action, the agent uses the 'old' action and, if available, delayed observation from $\tau$ timesteps before to update the belief $Q^s(t-\tau)$.
\begin{algorithm}
\caption{Active Inference for Mouse Point-and-Click}\label{alg:algorithm}
\begin{algorithmic}
\Require Parameters in~\cref{tab:parameters}
\State $t \gets 1$
\State $Q^s(1+i) \gets Q^s, \; \forall i \in \{-\tau, \dots, 0\}$
\State $a(1+i) \gets 0, \; \forall i \in \{-\tau,\dots, -1\}$
\While{Button not clicked}
    \State \textit{1. Planning (\cref{sec:action-selection}):}
    \State $\tilde{Q}^s(t-\tau) \gets Q^s_(t-\tau)$
    \For{$i \in \{-\tau, \dots, -1\}$}
        \State $\tilde{Q}^s(t+i+1) \gets$ Update $\tilde{Q}^s(t+i)$ using UKF and $a(t-i)$ (\cref{sec:belief-updates}, 1.)
    \EndFor
    \State $\pi_j \gets (a_{ij})_{ij}, \forall i \in \{1,\dots, N\}, j \in \{1, \dots, K\}; a_{ij} \sim \mathcal{U}(\underline{a},\overline{a})$ 
    \State $j \gets \argmin_{j \in \{1,\dots,K\}} G(\pi_j)$ (using $\tilde{Q}^s(t)$)
    \State $a(t) \gets \pi_j(1)$
    \State \textit{2. Step System:}
    \State $s(t+1) \gets f_\theta(s(t),a(t))$
    \State $o(t-\tau+1) \sim \mathcal{N}\left(g(s(t+1)),\sigma^p\right)$
    \State \textit{3. Update $Q^s$:}
    \State $Q^s(t-\tau+1) \gets$ Update $Q^s(t-\tau)$ via UKF and $a(t-\tau)$ (\cref{sec:belief-updates}, 1.)
    \If{$t > \tau$}
        \State $Q^s(t-\tau+1) \gets$ Update $Q^s(t-\tau+1)$ via VI and $o(t-\tau+1)$ (\cref{sec:belief-updates}, 2.)
    \EndIf
    \State $t \gets t +1$
\EndWhile
\end{algorithmic}
\end{algorithm}

\section{Experiment Details}\label{app:experiment-details}

\subsection{User Data Acquisition}\label{sec:data-acquisiton}
\begin{figure}
    \centering
    \begin{subfigure}[t]{0.5\linewidth}
        \centering  
        \includegraphics[width=0.9\linewidth]{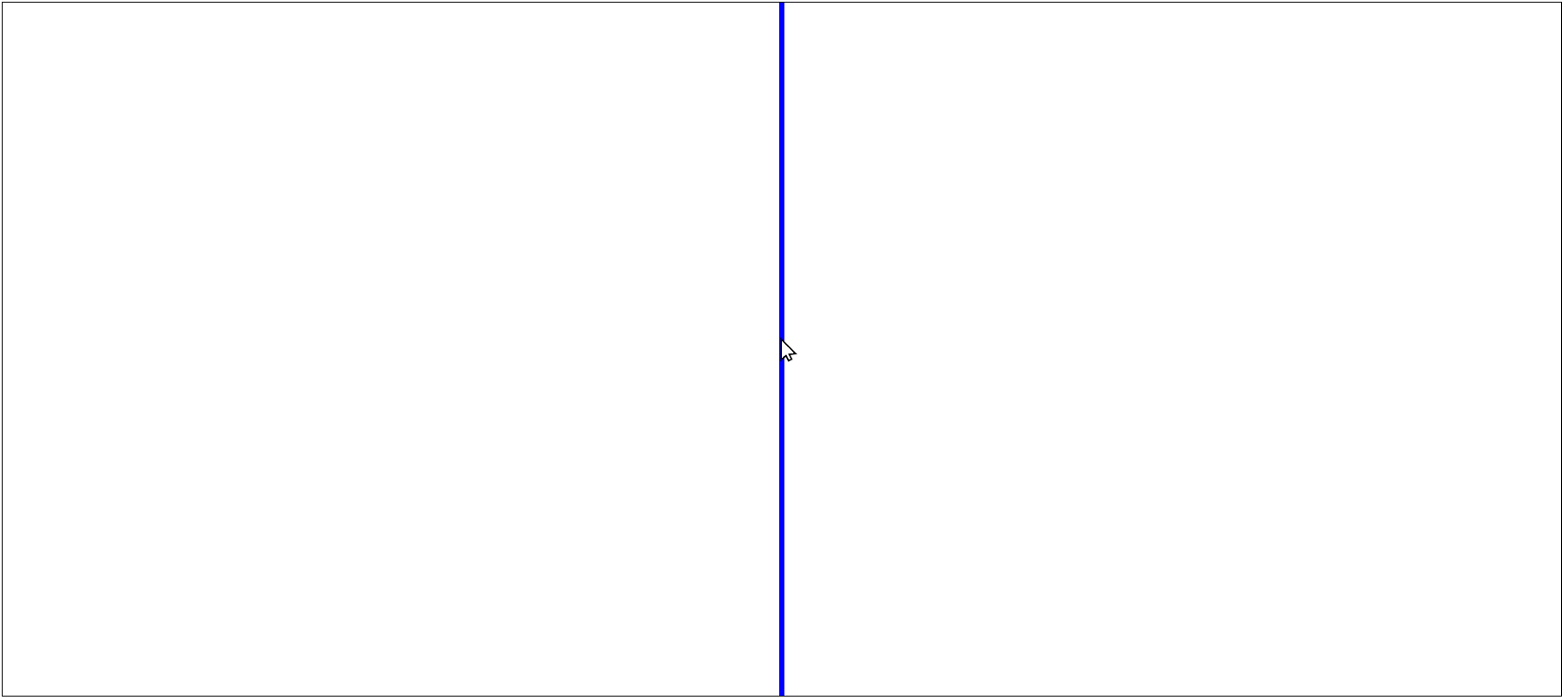}
        \caption{Start position}
    \end{subfigure}%
    ~ 
    \begin{subfigure}[t]{0.5\linewidth}
        \centering  
        \includegraphics[width=0.9\linewidth]{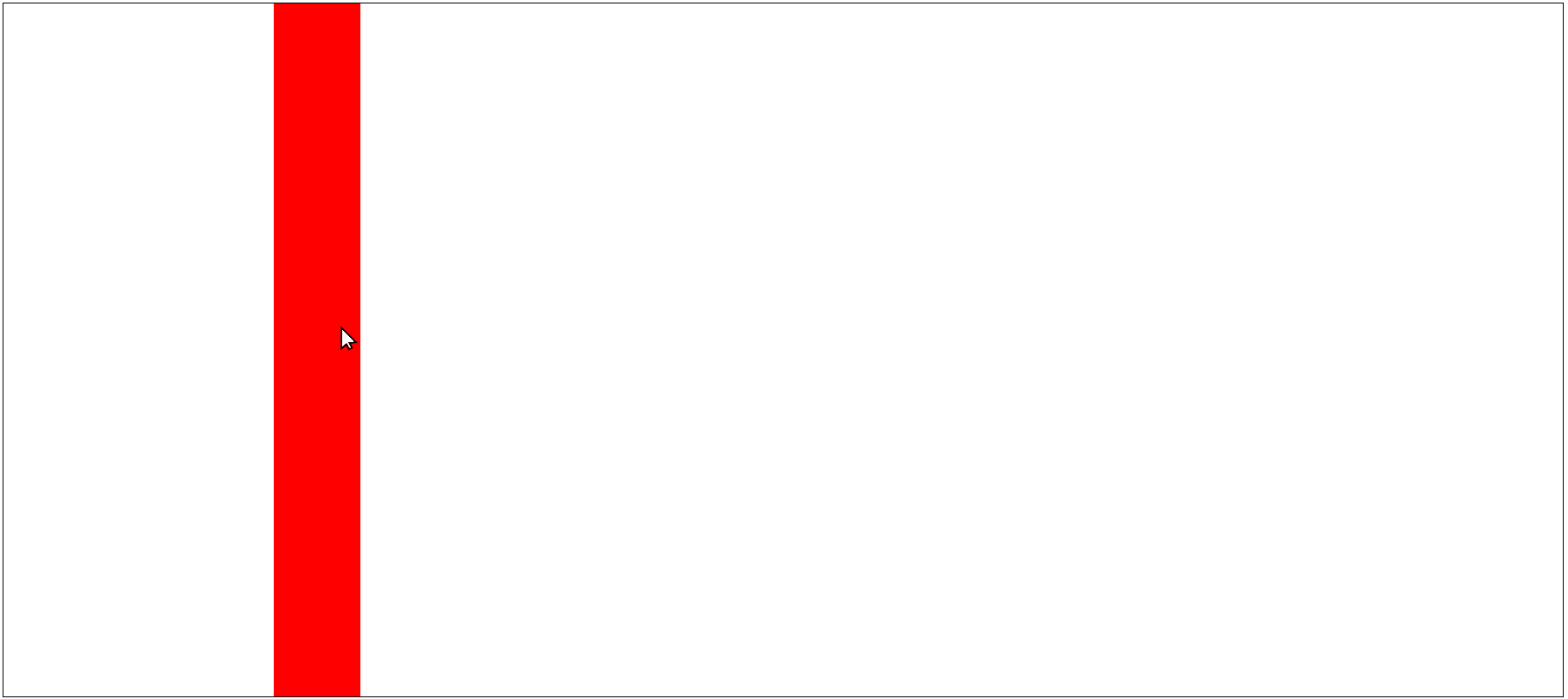}
        \caption{Target}
    \end{subfigure}
    \caption{The task commences by clicking on the designated start position, which is represented by the blue line. Subsequently, a red target is displayed, either to the right or the left. After clicking on the target, the cursor must be moved back to the initial position prior to starting the next trial.}
    \label{fig:1d-pointing-task}
\end{figure}

We gather movement data from a simple, PyScript-based application. 
The task is to move the mouse cursor from a start position to a target that appears in one of six random locations, either to the left or to the right of the start position at pixel 900.
The width of the canvas is set to 1800 pixels.
The distance between the start position and the target's centre is varied between 225, 537, 850 pixels, while its width is varied between 20, 60, 100 pixels, resulting in a total of 18 different targets.
The task is performed by six human users, hitting every target ten times in a random order, resulting in a total of 1080 movements.
We split the resulting data into single trials, and removed outliers that showed a movement time greater than three standard deviations away from the mean (a depiction of the task as well is shown in \cref{fig:1d-pointing-task}).
\cref{tab:targets} shows the 18 targets.

\FloatBarrier

\begin{table}[h!]
\centering
\begin{tabular}{c|c|c|c}
\textbf{Target Nr} & \textbf{Position} & \textbf{Width} & \textbf{ID} \\
\hline
\hline
1  & 675   & 20  & 3.61 \\
2  & 363   & 20  & 4.80 \\
3  & 50   & 20  & 5.44 \\
4  & 1125  & 20 & 3.61 \\
5  & 1438  & 20  & 4.80 \\
6  & 175  & 20 & 5.44 \\
\hline
7  & 675  & 60 & 2.25 \\
8  & 363   & 60 & 3.32 \\
9  & 50   & 60 & 3.92 \\
10 & 1125  & 60 & 2.25 \\
11 & 1438  & 60 & 3.32 \\
12 & 1750  & 60 & 3.92 \\
\hline
13 & 675   & 100 & 1.70 \\
14 & 363   & 100 & 2.67 \\
15 & 50    & 100 & 3.25 \\
16 & 1125  & 100& 1.70 \\
17 & 1438  & 100 & 2.67 \\
18 & 1750  & 100 & 3.25 \\
\end{tabular}
\caption{All targets of the pointing task with position, width in pixel, and index of difficulty.}
\label{tab:targets}
\end{table}

\section{Additional Results}\label{app:add_results}
\subsection{Changing Model Parameters}
\begin{figure}%
    \centering
    \begin{subfigure}[t]{0.5\linewidth}
        \centering  
        \includegraphics[height=.09\paperheight]{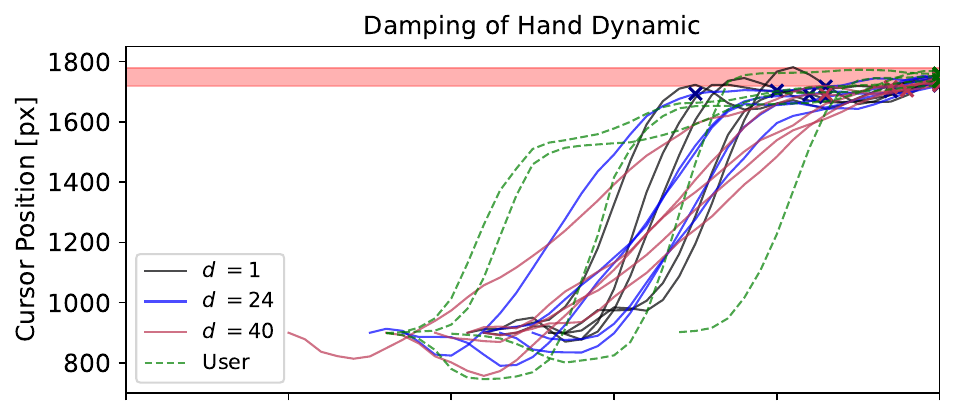}
    \end{subfigure}%
    \begin{subfigure}[t]{0.5\linewidth}
        \centering  
        \includegraphics[height=.09\paperheight]{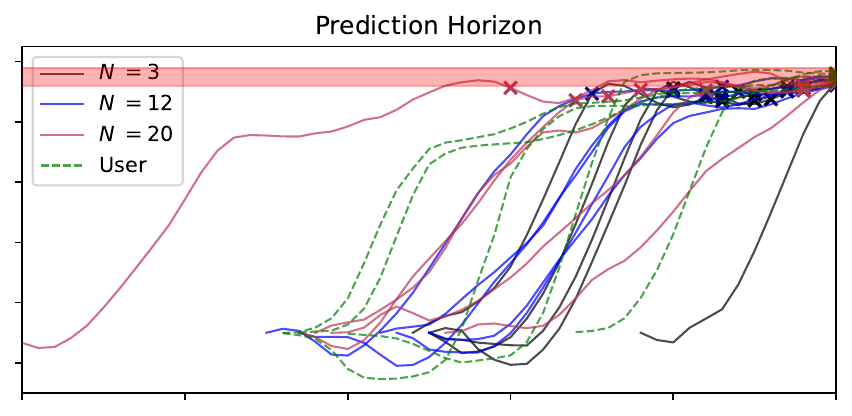}
    \end{subfigure}\\
    \begin{subfigure}[t]{0.5\linewidth}
        \centering  
        \includegraphics[height=.105\paperheight]{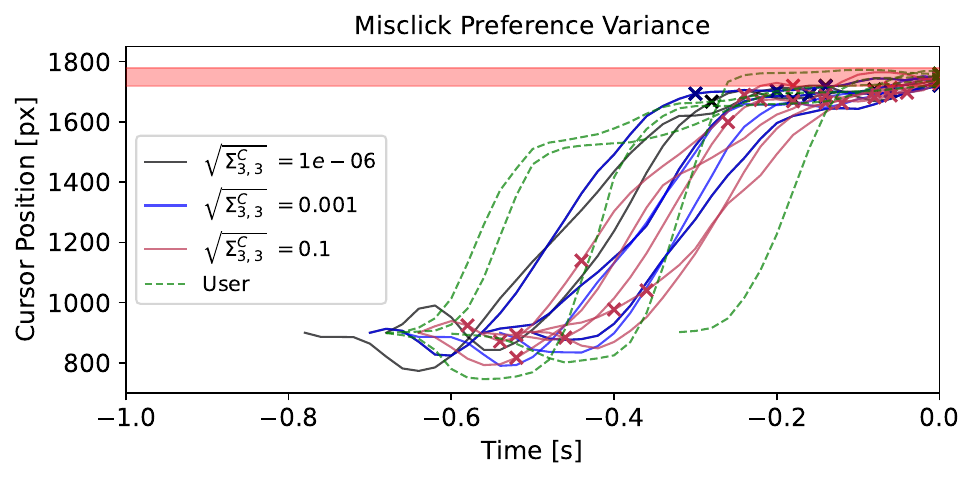}
    \end{subfigure}%
        \begin{subfigure}[t]{0.5\linewidth}
        \centering  
        \includegraphics[height=.105\paperheight]{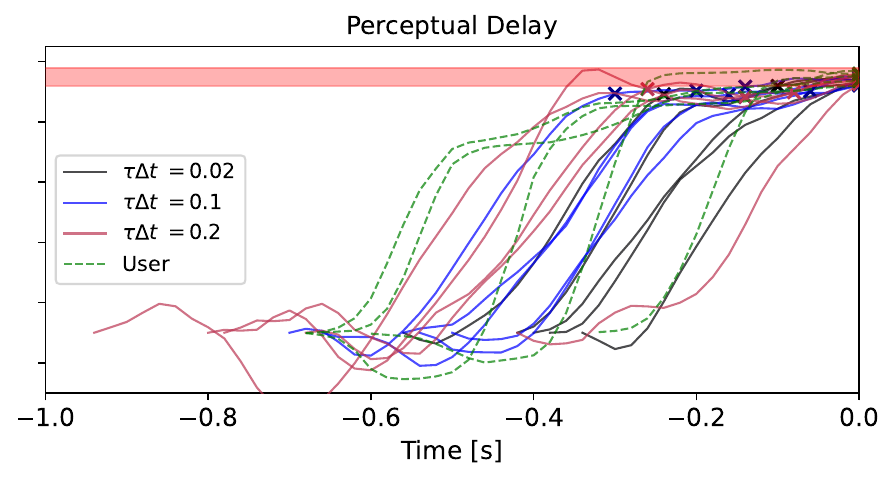}
    \end{subfigure}\\
    \caption{In each plot, only one parameter is changed while the others are fixed. The blue trajectories refer to the agent used in the remaining result section and the green trajectories are those of a human user. $t=0$ indicates the moment of a correct click. The crosses mark clicks and the red area the target.}

    \label{fig:param_sweep}
\end{figure}

Changing individual parameters can affect the point and click behaviour of the AIF agent heavily.
\cref{fig:param_sweep} shows trajectories to the same target ($T=1750$, $W=30$px), while adapting only one parameter.
Changing the damping parameter of the mouse cursor model (upper left) defines how steep the slope in the surge phase can be, with higher damping leading to a gentler slope (red trajectories, $d=40$).
The influence of the planning horizon (upper right) is harder to distinguish.
If the number of sampled plans during planning is not adjusted, the chance of sampling a good plan with a higher horizon is decreased.
Also, since the agent only applies the first action of the best plan, having a longer plan reduces the influence of this action on the expected free energy, which might lead to suboptimal control (long red trajectory).
Similarly, adjusting the preference of observing a misclick, i.e., changing $\sqrt{\Sigma^C_{3,3}}$, (lower left) has no clear effect on the resulting trajectories. 
Nonetheless, lower values lead to less misclicks (almost no black crosses, $\sqrt{\Sigma^C_{3,3}} = 10^{-6}$), while higher values can lead to many misclicks (many red crosses, $\sqrt{\Sigma^C_{3,3}} = 0.1$).
Increasing the agent's perceptual delay (lower right) increases the time to target in most cases (red lines, $0.2$s reaction time).
This is not surprising, since with longer delay the agent receives knowledge about the target position later and has a higher uncertainty about the cursor position after the surge phase (not shown).

\FloatBarrier

\end{document}